\documentclass[12pt]{article}
\usepackage{color}

\newcommand{\beq}{\begin{equation}}
\newcommand{\eeq}{\end{equation}}
\newcommand{\beqa}{\begin{eqnarray}}
\newcommand{\eeqa}{\end{eqnarray}}
\newcommand{\beqar}{\begin{eqnarray*}}
\newcommand{\eeqar}{\end{eqnarray*}}

\newcommand{\al}{\alpha}
\newcommand{\be}{\beta}

\def\non          {\nonumber}

\def\Tr           {\mbox{\rm Tr}\,}

\def\cd           {{\cdot}}
\def\ran          {\rangle}
\def\lan          {\langle}
\def\fsk    {k\!\!\!\!/\,}
\def\fsH    {H\!\!\!\!/\,}

\newcommand{\ga}{\gamma}

\newcommand{\lam}{\lambda}

\newcommand\bPsi{{\bar \Psi }}

\newcommand{\labell}[1]{\label{#1}} 
\newcommand{\reef}[1]{(\ref{#1})}

\newcommand\veps{\varepsilon}

\newcommand\cD{{\cal D}}

\newcommand\bu{\bar{u}}

\def\sst#1{{\scriptscriptstyle #1}}
\def\0{{\sst{(0)}}}
\def\1{{\sst{(1)}}}
\def\2{{\sst{(2)}}}
\def\3{{\sst{(3)}}}
\def\4{{\sst{(4)}}}
\def\5{{\sst{(5)}}}
\def\6{{\sst{(6)}}}
\def\7{{\sst{(7)}}}
\def\8{{\sst{(8)}}}

\newcommand{\dga}{\dot{\gamma}}
\newcommand{\dde}{\dot{\delta}}

\newcommand{\gab}{\bar{\gamma}}

\begin{document}
\baselineskip 18pt%
\begin{titlepage}
\vspace*{1mm}%
\hfill
\vbox{

    \halign{#\hfil         \cr
           } 
      }  
\vspace*{7mm}

\center{ {\bf \Large  More on Ramond-Ramond, SYM, WZ  couplings and their
corrections in IIA
}}\vspace*{1mm} \centerline{{\Large {\bf  }}}
\begin{center}
{Ehsan Hatefi \small $^{1,2,3,4}$}

\vspace*{0.6cm}{ {\it
{\small $^{1}$International Centre for Theoretical Physics, Strada Costiera 11, Trieste, Italy},
\vskip.1in
{\small $^{2}$ Centre for Research in String Theory, School of Physics and Astronomy,
Queen Mary University of London,Mile End Road, London E1 4NS, UK},
\vskip.06in

{\small $^{3}$National Institute for Theoretical Physics ,
School of Physics and Centre for Theoretical Physics,University of the Witwatersrand, Wits, 2050, South Africa}
\vskip.06in
 {\small $^{4}$E-mails:ehsan.hatefi@cern.ch,ehsanhatefi@gmail.com}

}}
\vspace*{0.1cm}
\vspace*{.1cm}
\end{center}
\begin{center}{\bf Abstract}\end{center}
\begin{quote}

We obtain the closed form of the correlation function of one current and four spin operators (with different chiralities ) in type IIA superstring theory. The complete  form of the S-matrix of one gauge , two fermions (with different chiralities) and one closed string Ramond-Ramond  for all order $\alpha'$ type IIA has been explored. Moreover, we make use of  a different  gauge fixing  to be able to find all the closed forms of   $<V_{C} V_{A}V_{\bar\psi}V_{\psi} >$ correlators. An infinite number of  $t,s-$channel fermion poles of this S-matrix  in the field theory of IIA is generated. Unlike the
the closed form of the correlators of the same amplitude of IIB, for various $p,n$ cases  in type IIA we do have different  double poles in $t,(t+s+u)$ and in $s,(t+s+u)$ channels and we produced them. We also find new Wess-Zumino couplings of IIA  with their third order $\alpha' $ corrections as well as a new form of higher derivative corrections  (with different coefficient from its IIB one ) to SYM couplings at third order of $\alpha' $.  Using them, we are able to produce a $(t+s+u)$ channel scalar pole of the $<V_{C} V_{A}V_{\bar\psi}V_{\psi} >$ amplitude for $p+2=n$ case. Finally we make some comments on $\alpha'$ corrections to IIA Superstring theory.

\end{quote}
\end{titlepage}

\section{Introduction}

It is known that the fundamental objects, the so called D$_p$-branes  ($p$ is spatial dimension of brane) \cite{Polchinski:1995mt},\cite{Witten:1995im},\cite{Polchinski:1996na}  could play the most important rule in both type IIA and IIB superstring theories as well as in Ads/CFT correspondence. Let us start mentioning several crucial papers on superstring theories \cite{Witten:2013tpa,Witten:2013cia,D'Hoker:2013eea,Bjerrum-Bohr:2014qwa} and refer to some pioneering papers that pointed out either mathematical structures or dealt with various symmetries of the scattering amplitudes/ gauge theories \cite{Arkani-Hamed:2013jha,Arkani-Hamed:2013kca,ArkaniHamed:2012nw}. To reveal the dynamics of a brane one needs to  consider diverse  transitions of open/closed strings, where several processes have been very well realized in \cite{Ademollo:1974fc}.
 \vskip.1in

 Given several reasons, we are not interested in applying dualities, however,  it is always nice to have understandings of string dualities and clearly reveal most of the dual descriptions of IIA (IIB) theories that to our knowledge originated from \cite{Polchinski:1996nb}.\footnote{ As an instance of dual prescriptions , one might point out the mixed system of  $D0/D4$ with some particular applications  \cite{Hatefi:2012sy}.}

  \vskip.1in

 In this paper we try to find out the effective actions of type IIA for all BPS branes\footnote{They accompany RR charge as well.} of two fermions (with different chirality ), one closed string Ramond-Ramond (RR) and a massless gauge field in the world  volume space by making use of superstring scattering from different D-branes. One could see some part of those actions from the point of view of the supergravity \cite{Hatefi:2012bp}, for which ADM formalism could be applied to IIB and the realization of both Ads (ds) space brane world is achieved   \cite{Hatefi:2012bp}.\footnote{ In  \cite{Polchinski:1994fq} different boundary conditions are applied to actually observe the appearance of D-branes in flat de-compactification space.}

\vskip 0.2in

 Concerning the complete form of IIA S-matrix, we are able to first of all find out the third order of $\alpha'$ higher derivative corrections to SYM couplings and secondly distinguish $\alpha'$ corrections from type IIB.  Our computations show that the coefficients of the corrections should be different in type IIA while in  \cite{Hatefi:2014saa}  we have proven that even the the general structure of IIA corrections of two fermions two scalars  is different from its IIB one.

\vskip.2in

 In order to talk about the dynamical facet of various branes, we should focus on the proper effective actions. The bosonic effective actions in the presence of different configurations are addressed in Myers paper \cite{Myers:1999ps} with some clarifications about all order higher derivative corrections added in \cite{Hatefi:2012zh}. On the other hand with our knowledge, the super symmetrized actions of those bosonic actions  are not  yet completely derived, however, let us address an original work in this topic \cite{Howe:2006rv}. For a single  bosonic D$_p$-brane and for its super symmetrized action we point out to  \cite{Leigh:1989jq} and  \cite{Cederwall:1996pv} appropriately.
 \vskip .2in

In order to avoid explaining all the details of the description of the world volume brane dynamics (Myers terms, the Chern-Simons, Wess-Zumino (WZ) and Born-Infeld action) we just refer to the basic references \cite{Hatefi:2012zh,Tseytlin:1999dj,Tseytlin:1997csa,Hatefi:2010ik}. Three standard approaches to the effective field theory of  either Myers terms or  Taylor expansion / Pull-back methods including the method for exploring all order  $\alpha'$ higher derivative corrections are explicitly (within the complete details) explained in \cite{Hatefi:2012wj}. Nevertheless, we argued in \cite{Hatefi:2012zh} that some other methods are needed to be able to construct new couplings of either BPS \cite{Hatefi:2012zh}, \cite{Hatefi:2013eia} or non-supersymmetric branes \cite{Hatefi:2013yxa,Hatefi:2013mwa,Hatefi:2012cp,Garousi:2007fk,Hatefi:2008ab}.

\vskip 0.2in

Some great results due to the efforts \cite{Koerber:2002zb},\cite{Keurentjes:2004tu}, \cite{Denef:2000rj} have been already come out. Our first goal of this paper is just to employ the direct scattering amplitude process to see whether or  not one could extract more information on the general structure of the effective actions in IIA and the second goal is to describe in an efficient way the dynamical aspects of branes. We introduce
\cite{Hashimoto:1996bf} to notice the scattering of BPS branes or to provide more applications  to the BPS/non-BPS branes \cite{Hashimoto:1996kf,Lambert:2003zr,Dudas:2001wd,Antoniadis:1999xk,deAlwis:2013gka}. Eventually having worked new WZ and Myers terms \cite{Hatefi:2010ik,Hatefi:2012cp,Garousi:2007fk,Hatefi:2008ab,Ferrari:2013pi,Hatefi:2012ve,Hatefi:2012rx}, we can definitely get to some of the corrections of string theory \cite{Boels:2010bv}.

\vskip 0.2in

Given the fact that there is no derivation for AdS/CFT correspondence, it is always good to have some theoretical methods at hand to get to $\alpha'$ higher derivative corrections of IIA superstring theory. Indeed there is a very direct correspondence both in Ads/CFT and in string theory between a closed and an open string so it should be an  inevitable task to have embedded the corrections in a compact form. Thus if we deal with open-closed amplitudes of type II, we may hope to shed light on  various features of corrections, dualities and some other works. As a specific example, we have obtained some  WZ  couplings  involving their $\alpha'$ higher derivative corrections and in particular it was shown how those corrections might be taken to interpret  the $N^3$ entropy of M5 brane \cite{Hatefi:2012sy} so that it could be realized as a matter of dissolving branes with lower dimensions inside the higher dimensions. It also happens for
  $D(-1)/D3$ system in which by applying  those $\alpha'$  corrections we could interpret  $N^2$ entropy relation.  $\alpha'$ corrections, new  WZ couplings are also applied to flux vacua, M or F-theory frameworks \cite{Maxfield:2013wka,McOrist:2012yc,Vafa:1996xn}.\footnote{For other applications we highlight  two different conjectures on the  quantum effects of the BPS strings \cite{Park:2007mc} and  \cite{Hatefi:2012rx}, holding for all order $\alpha'$ higher derivative corrections to BPS/non-BPS branes  of bosonic (IIA,IIB) and fermionic amplitudes of type IIB.}

\vskip 0.2in

Here is the organization of the paper.

We first apply conformal field theory techniques (CFT) to discover the  compact/closed form of the correlation function of four spin operators with different chirality and one current  $<V_{C}V_{\bar\psi}^{\dga}V_{\psi}^{\dde} V_{A}>$ in ten dimension of space time in type IIA superstring theory. The entire form of the S-matrix of one closed string RR, two fermion fields with different chirality and one gauge field of IIA   with all order $\alpha'$ closed form have also  been explored. Let us highlight the fact that due to several differences , we can not obtain this S-matrix from its IIB one.

Because first of all unlike type IIB amplitude there is no any u-channel scalar/gauge pole left over.
Secondly, we will obtain various double poles in $t,(t+s+u)$ and  in $s,(t+s+u)$ channels of closed form of $<V_{C}V_{\bar\psi} V_{\psi} V_{A}>$ of IIA but there is no any double pole in IIB and  the presence of these double poles can not be confirmed by duality transformation at all.
We also generate all infinite massless fermion poles of   $t,s$-channels of type IIA  for diverse $p,n$ ($n$ is the rank of RR field strength) cases and the extensions of several new Wess-Zumino couplings have been confirmed.

\vskip 0.1in

Most importantly their all order $\alpha'$ higher derivative corrections (related to contact interactions out of fermion poles) without any ambiguity are found out.
 Thus our computations are different from IIB, as one could clearly confirm this sharp conclusion by comparing  both S-matrix computations  of  all two different theory.  We also show that there are no corrections to two fermion two gauge couplings. Having obtained the entire form of the S-matrix, we are able to discover several new Wess-Zumino couplings
with their third order $\alpha' $ correction and note that those couplings can not be found by having IIB form of the S-matrix. It is just the first  $(t+s+u)-$ channel scalar pole for $p+2=n$ case dictating us that the $\alpha'^3$  corrections of two fermions, one scalar one gauge of IIA have to be accompanied with the same corrections but with different coefficients of their corrections in IIB. We take this fact as an advantage of carrying out the direct CFT rather than applying duality transformation.\footnote{ More promising explanations are clarified in  \cite{Hatefi:2012zh}
  and we have shown that one is not able to derive the  entire result of  $<V_{C}V_{\phi} V_AV_A>$ from   $<V_{C}V_A V_AV_A>$ by using T-duality transformation to \cite{Hatefi:2010ik}.}

\section{  The Complete form of   $CA\bar\Psi\Psi$  S-matrix in IIA}

To get to the complete and closed form of the amplitude of one closed string RR, two fermions with different chirality and one gauge field in the world volume of BPS branes of IIA, one must use the direct  CFT techniques. The reason for this stark conclusion will be clarified in the next sections.  We try to address some of the higher point stable BPS functions
\cite{Hatefi:2010ik,Hatefi:2012ve,Hatefi:2012rx,Bilal:2001hb,Barreiro:2012aw,Barreiro:2013dpa,Kennedy:1999nn,Chandia:2003sh} and also emphasize on various  non-BPS \cite{Hatefi:2012wj,Hatefi:2013mwa} computations. It is obvious that non-BPS S-matrices can not be derived by making use of any dualities (see \cite{Hatefi:2012cp}). The fermion vertex operators of type IIA are the same as type IIB, however,  their chirality in IIA must be changed for which  for the completeness we just write them down in the footnote. \footnote{

\beqa
V_{A}^{(-2)}(y) &=&e^{-2\phi(y)}V_{A}^{(0)}(y),
\nonumber\\
V_{A}^{(0)}(x) &=& \xi_{a}\bigg(\partial
X^a(x)+\alpha' ik\cd\psi\psi^a(x)\bigg)e^{\alpha' ik\cd X(x)},
\nonumber\\
V_{\bPsi}^{(-1/2)}(x)&=&\bu^{\dga}e^{-\phi(x)/2}S_{\dga}(x)\,e^{\alpha'iq.X(x)} \nonumber\\
V_{\Psi}^{(-1/2)}(x)&=&u^{\dde}e^{-\phi(x)/2}S_{\dde}(x)\,e^{\alpha'iq.X(x)} \nonumber\\
V_{C}^{(-\frac{1}{2},-\frac{1}{2})}(z,\bar{z})&=&(P_{-}\fsH_{(n)}M_p)^{\al\be}e^{-\phi(z)/2}
S_{\al}(z)e^{i\frac{\alpha'}{2}p\cd X(z)}e^{-\phi(\bar{z})/2} S_{\be}(\bar{z})
e^{i\frac{\alpha'}{2}p\cd D \cd X(\bar{z})}.
\label{d4Vs}
\eeqa}
\vskip.2in

All on-shell Conditions,  the definitions of RR's field strength and charge conjugation in type IIB and type IIA  are given accordingly  in  \cite{Hatefi:2014saa} and \cite{Hatefi:2013eia,Hatefi:2013hca}. To work with holomorphic functions, we apply the doubling tricks to deal with the usual correlation functions for bosonic and fermionic world sheet fields.\footnote{
\begin{eqnarray}
\lan X^{\mu}(z)X^{\nu}(w)\ran & = & -\frac{\alpha'}{2}\eta^{\mu\nu}\log(z-w) , \non \\
\lan \psi^{\mu}(z)\psi^{\nu}(w) \ran & = & -\frac{\alpha'}{2}\eta^{\mu\nu}(z-w)^{-1} \ ,\non \\
\lan\phi(z)\phi(w)\ran & = & -\log(z-w) \ .
\labell{prop}\end{eqnarray} To see  more standard notations and further details we suggest \cite{Hatefi:2012wj,Hatefi:2012ve}.}

\vskip.3in

  Since the correlation function of two fermions and one gauge field in type IIA is exactly the same as type IIB, the S-matrices are also the same as they had been computed in \cite{Polchinski:1998aa}. Hence, one could easily show that by taking into account these definitions, \footnote{
 $(2\pi\alpha' T_p)\Tr(\bar\psi\ga^{a} D_{a}\psi), \quad D^a\psi=\partial^a\psi-i[A^a,\psi]$}
$<V_{A}V_{\bar\psi}V_{\psi} >$ S-matrix could be generated in the same manner in the field theory of IIA as well as IIB,however, for mixed higher point functions the work gets complicated as we comment on it.

\vskip.5in

We put RR in $(-1)$ picture $(V_{RR}^{(-\frac{1}{2},-\frac{1}{2})}(z,\bar{z}) )$, fermions in
$(V_{\bar\psi}^{(-1/2)}{(x_{2})}V_{\psi}^{(-1/2)}{(x_{3})})$
(their standard pictures) and since the total ghost charge must be $(-2)$, we then have to put the gauge field just in zero picture $( V_{A}^{(0)}{(x_{1})})$.
We find the complete form of this S-matrix in IIA and we show that this S-matrix is  different from IIB thus all closed form of the corrections of type IIB can not be used for IIA and vice versa.

\vskip.1in

For the first part of the S-matrix we need to work with the correlation function of four spin operators with different chirality \cite{Friedan:1985ge}.\footnote{

  $<S_{\alpha}(z_4)S_{\beta}(z_5)S^{\dga}(z_2)S^{\dde}(z_3)>= \bigg(\frac{x_{45}x_{23}}{x_{42}x_{43}x_{52}x_{53}}\bigg)^{1/4}
  \bigg[\frac{C_{\alpha}^{\dde}C_{\beta}^{\dga}} {x_{43}x_{52}}-\frac{C_{\alpha}^{\dga}C_{\beta}^{\dde}} {x_{42}x_{53}}+\frac{1}{2}\frac{(\gamma^\mu C)_{\alpha\beta}\,(\gab_\mu C)^{\dga\dde}} {x_{45}x_{23}}\bigg]$} Having set that, we find the first part of the mixed amplitude of IIA as follows

\beqa {\cal A}_{1}&\sim &
 \int
 dx_{1}dx_{2}dx_{3}dx_{4} dx_{5}\,
(P_{-}\fsH_{(n)}M_p)^{\alpha\beta}\xi_{1a} \bu ^{\dga} u^{\dde}  (x_{23}x_{24}x_{25}x_{34}x_{35}x_{45})^{-1/4}
\nonumber\\&&\times \bigg(\frac{x_{45}x_{23}}{x_{42}x_{43}x_{52}x_{53}}\bigg)^{1/4}
  \bigg[\frac{C_{\alpha}^{\dde}C_{\beta}^{\dga}} {x_{43}x_{52}}-\frac{C_{\alpha}^{\dga}C_{\beta}^{\dde}} {x_{42}x_{53}}+\frac{1}{2}\frac{(\gamma^\mu C)_{\alpha\beta}\,(\gab_\mu C)^{\dga\dde}} {x_{45}x_{23}}\bigg] I_1
 \Tr(\lam_1\lam_2\lam_3),\labell{125}\eeqa
  with
\beqa
I_1 &=& \bigg[ik_2^a \bigg(\frac{x_{42}}{x_{14}x_{12}}+ \frac{x_{52}}{x_{15}x_{12}}\bigg)+ik_3^a \bigg(\frac{x_{43}}{x_{14}x_{13}}+ \frac{x_{53}}{x_{15}x_{13}}\bigg)\bigg]|x_{12}|^{\alpha'^2 k_1.k_2}|x_{13}|^{\alpha'^2 k_1.k_3}|x_{14}x_{15}|^{\frac{\alpha'^2}{2} k_1.p}
\nonumber\\&&\times
 |x_{23}|^{\alpha'^2 k_2.k_3}|
x_{24}x_{25}|^{\frac{\alpha'^2}{2} k_2.p}
|x_{34}x_{35}|^{\frac{\alpha'^2}{2} k_3.p}|x_{45}|^{\frac{\alpha'^2}{4}p.D.p},
\nonumber\eeqa

 In above we wrote the S-matrix in such a way that it clearly shows the the property of SL(2,R) invariance of the amplitude \footnote{ With the following change of variables
$x_{ij}=x_i-x_j, x_4=z=x+iy, x_5=\bar z=x-iy$ \beqa
s&=&-\frac{\alpha'}{2}(k_1+k_3)^2, \quad t=-\frac{\alpha'}{2}(k_1+k_2)^2, \quad u=-\frac{\alpha'}{2}(k_3+k_2)^2,
\nonumber
\eeqa}.

\vskip.3in

In order to explore the closed and complete form of the amplitude with their $\alpha'$ corrections , one must take integrations on the location of RR closed string, and it means that we have to get rid of the position of open strings by having applied a very particular gauge fixing as $(x_1=0,x_2=1,x_3=\infty)$. If we do apply that gauge fixing, then we get to the following :

\beqa {\cal A}_{1}^{C A\bar\psi \psi}& \sim & (P_{-}\fsH_{(n)}M_p)^{\alpha\beta}\xi_{1a} \bu_{\dga} u_{\dde} \int\int
 dz d\bar z |z|^{2t+2s}|1-z|^{2t+2u-1} (z-\bar z)^{-2(t+s+u)}\Tr(\lam_1\lam_2\lam_3)
\nonumber\\&&\times
\bigg[\frac{C_{\alpha}^{\dde}C_{\beta}^{\dga}} {1-\bar z}-\frac{C_{\alpha}^{\dga}C_{\beta}^{\dde}} {1-z}-\frac{1}{2}\frac{(\gamma^\mu C)_{\alpha\beta}\,(\gab_\mu C)^{\dga\dde}} {z-\bar z}\bigg]
 \bigg(2ik_2^a-\frac{(z+\bar z)(ik_2^a+ik_3^a)}{|z|^2}\bigg)
\labell{amp3q},\eeqa

\vskip 0.1in

 We need to deal with several important results for the integrals on the upper half plane. For the fourth term above we need to employ the results for the integrals taken in \cite{Hatefi:2012wj} and for the other integrals in the above amplitude we use \cite{Fotopoulos:2001pt} . Thus one can write down the final result for the first part of the amplitude as below

\beqa
{\cal A}_{1}^{C A\bar\psi \psi}& \sim& (P_{-}\fsH_{(n)}M_p)^{\alpha\beta}\xi_{1a} \bu ^{\dga} u ^{\dde}
  \bigg\{ik_2^a \bigg[ sL_1 (C_{\alpha}^{\dde}C_{\beta}^{\dga}+C_{\alpha}^{\dga}C_{\beta}^{\dde})-2sL_2(C_{\alpha}^{\dde}C_{\beta}^{\dga}-C_{\alpha}^{\dga}C_{\beta}^{\dde})
  \bigg]\nonumber\\&&-ik_3^a\bigg[tL_1 (C_{\alpha}^{\dde}C_{\beta}^{\dga}+C_{\alpha}^{\dga}C_{\beta}^{\dde})-2tL_2(C_{\alpha}^{\dde}C_{\beta}^{\dga}-C_{\alpha}^{\dga}C_{\beta}^{\dde})
  \bigg]\nonumber\\&&+[ik_2^a s-ik_3^at](\gamma^\mu C)_{\alpha \beta}(\bar\gamma_\mu C)_{\dga\dde}L_3 \bigg\} \Tr(\lam_1\lam_2\lam_3)
\labell{amp3q},\eeqa

\vskip.2in

with
\beqa
L_1&=&(2)^{-2(t+s+u)+1}\pi{\frac{\Gamma(-u+\frac{3}{2})
\Gamma(-s)\Gamma(-t)\Gamma(-t-s-u+1)}
{\Gamma(-u-t+\frac{3}{2})\Gamma(-t-s+1)\Gamma(-s-u+\frac{3}{2})}},\nonumber\\
L_2&=&(2)^{-2(t+s+u)}\pi{\frac{\Gamma(-u+1)
\Gamma(-s+\frac{1}{2})\Gamma(-t+\frac{1}{2})\Gamma(-t-s-u+\frac{1}{2})}
{\Gamma(-u-t+\frac{3}{2})\Gamma(-t-s+1)\Gamma(-s-u+\frac{3}{2})}},\nonumber\\
L_3&=&(2)^{-2(t+s+u)-1}\pi{\frac{\Gamma(-u+\frac{1}{2})
\Gamma(-s)\Gamma(-t)\Gamma(-t-s-u)}
{\Gamma(-u-t+\frac{1}{2})\Gamma(-t-s+1)\Gamma(-s-u+\frac{1}{2})}},\nonumber\\
\label{Ls}
\eeqa

\vskip.1in

 Obviously as we expected the amplitude is antisymmetric  with respect to interchanging the fermions. All the terms including $L_2$ are related to contact interactions, on the other hand the terms carrying the coefficients of $-sL_1$ ($-tL_1$) have an infinite singularities in $t(s)$ channels.

 In the closed form of the correlators of one RR, two fermions and one gauge field of type IIB, we did not have any double pole while here as it is clear from the closed form of the first part of the amplitude in type IIA, we do have various double poles. Basically  $-sL_3$ ($-tL_3$) imposed us the fact that those parts of the S-matrix do include double poles in $t$ and $(t+s+u)$ ($s$ and $t+s+u$)  channels accordingly.
 Of course if  $\mu$ gets world volume (transverse) index then we will have gauge for $n=p$ case (scalar for $n=p+2$ case) poles appropriately, where the expansion is just low energy expansion and all singularities/contact interactions would be generated by having sent all three Mandelstam variables to zero \cite{Hatefi:2010ik}.

 \vskip.1in

Now we get to derive the second part of the S-matrix in type IIA.  In order to proceed one has to know the closed form of the correlation function of four spin operators (with different chirality in ten dimensions) and one current. Indeed the  current which consists of two fermion fields in one position  comes from the second part of the vertex of the massless gauge field in zero picture.

 The method for obtaining this correlation function has been completely explained in \cite{Hatefi:2013eia} and it has been mentioned how to build various combinations of the Gamma matrices where  we have also used the equations/identities explained in   \cite{Hatefi:2014saa}.   Basically one has to first take the OPE of the current with one spin operator (the result is just spin operator ) and then replace it inside the main correlation function (four spin operators) and carry out the same approaches for the other cases. Eventually re-combine different combinations of various symmetric and antisymmetric gamma matrices. Given the comprehensive explanations in \cite{Hatefi:2014saa,Hatefi:2013hca} , one can find out the entire and closed form of the correlation function of four spin operators with different chirality with one current in ten dimensions of space time in type IIA as follows:

\vskip.1in

 \beqa
 &&<\psi^a\psi^i(z_1) S^{\dga}(z_2)S^{\dde}(z_3) S_{\alpha}(z_4)S_{\beta}(z_5)>=\frac{(x_{45} x_{23})^{-3/4}(x_{42} x_{43} x_{52} x_{53})^{-1/4}}{(x_{14} x_{15} x_{12} x_{13})} \nonumber\\&&\times   \bigg[-\frac{1}{2}(\gamma^a \gab^i C)_\alpha{}^{\dga} C_\beta^{\dde}
   \frac{x_{13}x_{15}x_{45}x_{23}}{x_{53}}
  +\frac{1}{2}(\gamma^a\gab^i C)_\alpha{}^{\dde}C_\beta^{\dga}\frac{x_{12} x_{15}x_{45}x_{23}}{x_{52}} +
  \frac{1}{2}(\gamma^a \gab^i C)_\beta{}^{\dga}C_\alpha^{\dde}\nonumber\\&&\times\frac{x_{14}x_{13}x_{45}x_{23}}{x_{43}}
   -\frac{1}{2}(\gamma^a \gab^i C)_\beta{}^{\dde}C_\alpha^{\dga}
  \frac{x_{14}x_{12}x_{45}x_{23}}{x_{42}}
  -\frac{1}{2}(\gamma^a C)_{\alpha\beta}(\gab^i C)^{\dga\dde}x_{14}x_{13}x_{25}\nonumber\\&&+
  \frac{1}{2}(\gamma^i C)_{\alpha\beta}(\gab^a\,C)^{\dga\dde}x_{14}x_{12}x_{53}
  -\frac{1}{4}(\gamma^a\gab^\lambda C)_\alpha{}^{\dga} (\gamma^i \gab_\lambda C)_\beta{}^{\dde}x_{13}x_{12}x_{45}\nonumber\\&&
  -\frac{1}{4}(\gamma^a \gab^\lambda C)_\alpha{}^{\dde}(\gamma^i \gab_\lambda C)_\beta{}^{\dga}x_{12}x_{13}x_{45}
  +\frac{1}{4}(\gab^a \gamma^i \gab^\lambda C)^{\dga\dde}(\gamma_\lambda C)_{\alpha\beta}x_{14}x_{15}x_{23}\bigg]
   \eeqa

 Hence, one can write down the second part of the S-matrix as below:

 \beqa {\cal A}_{2}^{C A\bar\psi \psi}& \sim &  \int
 dx_{1}dx_{2}dx_{3}dx_{4} dx_{5}\,
(P_{-}\fsH_{(n)}M_p)^{\alpha\beta}\xi_{1a} (2ik_{1b})\bu^{\dga} u^{\dde} ( x_{23}x_{24}x_{25}x_{34}x_{35} x_{45})^{-1/4} \nonumber\\&&\times
 <:\psi^b\psi^a(x_1):S_{\dga}(x_2):S_{\dde}(x_3):S_{\alpha}(x_4):S_{\beta}(x_5):>
 I
 \Tr(\lam_1\lam_2\lam_3),\labell{1289}\eeqa
with\footnote{
\beqa
I=|x_{12}|^{\alpha'^2 k_1.k_2}|x_{13}|^{\alpha'^2 k_1.k_3}|x_{14}x_{15}|^{\frac{\alpha'^2}{2} k_1.p}|x_{23}|^{\alpha'^2 k_2.k_3}|
x_{24}x_{25}|^{\frac{\alpha'^2}{2} k_2.p}
|x_{34}x_{35}|^{\frac{\alpha'^2}{2} k_3.p}|x_{45}|^{\frac{\alpha'^2}{4}p.D.p},
\nonumber\eeqa}

 \vskip.2in

Now by replacing the closed form of the $<:\psi^b\psi^a(x_1):S_{\dga}(x_2):S_{\dde}(x_3):S_{\alpha}(x_4):S_{\beta}(x_5):>$ correlator, gauge fixed it as we did for the first part of the amplitude and taking the integrations on the position of the RR , we get the final result of the amplitude in type IIA as below:

 \vskip.2in

\beqa {\cal A}_{2}^{C A \bar\psi \psi}  & \sim &  (P_{-}\fsH_{(n)}M_p)^{\alpha\beta}\xi_{1a}(2ik_{1b}) \bu^{\dga} u^{\dde}  \bigg({\cal A}_{21}+{\cal A}_{22}+{\cal A}_{23}+{\cal A}_{24}+{\cal A}_{25}\bigg)
\Tr(\lam_1\lam_2\lam_3),\nonumber\eeqa

such that
\beqa {\cal A}_{21} &\sim&\bigg\{-\frac{1}{2}(\gamma^b \gab^a C)_\alpha{}^{\dga} C_\beta^{\dde}
\bigg[L_5+\frac{s}{2} L_6\bigg]+\frac{1}{2}(\gamma^b \gab^a C)_\beta{}^{\dga}C_\alpha^{\dde}
\bigg[L_5-\frac{s}{2} L_6\bigg]\bigg\}\frac{1}{(-s-u+\frac{1}{2})}
\nonumber\\
 {\cal A}_{22} &\sim&\bigg\{\frac{1}{2}(\gamma^b\gab^a C)_\alpha{}^{\dde}C_\beta^{\dga}
\bigg[-L_5+\frac{t}{2} L_6\bigg]-\frac{1}{2}(\gamma^b \gab^a C)_\beta{}^{\dde}C_\alpha^{\dga}
\bigg[-L_5-\frac{t}{2} L_6\bigg]\bigg\}\frac{1}{(-t-u+\frac{1}{2})}
\nonumber\\
{\cal A}_{23} &\sim&-\frac{1}{2}(\gamma^b C)_{\alpha\beta}(\gab^a C)^{\dga\dde}
\bigg[sL_3+\frac{1}{2} L_4\bigg]-\frac{1}{2}(\gamma^a C)_{\alpha\beta}(\gab^b\,C)^{\dga\dde}
\bigg[-tL_3+\frac{1}{2} L_4\bigg]
\nonumber\\
{\cal A}_{24} &\sim& L_4\bigg[-\frac{1}{4}(\gamma^b\gab^\lambda C)_\alpha{}^{\dga} (\gamma^a \gab_\lambda C)_\beta{}^{\dde}
  -\frac{1}{4}(\gamma^b \gab^\lambda C)_\alpha{}^{\dde}(\gamma^a \gab_\lambda C)_\beta{}^{\dga} \bigg]
\nonumber\\
{\cal A}_{25} &\sim& -(-t-s) L_3\bigg[\frac{1}{4}(\gab^b \gamma^a \gab^\lambda C)^{\dga\dde}(\gamma_\lambda C)_{\alpha\beta}\bigg]
\labell{ampc}
\eeqa

with the following definitions
\beqa
L_4&=&(2)^{-2(t+s+u)}\pi{\frac{\Gamma(-u)
\Gamma(-s+\frac{1}{2})\Gamma(-t+\frac{1}{2})\Gamma(-t-s-u+\frac{1}{2})}
{\Gamma(-u-t+\frac{1}{2})\Gamma(-t-s+1)\Gamma(-s-u+\frac{1}{2})}}
\label{Ls24}
\eeqa

such that $L_5=-u L_4$ and does not involve any singularity and also $L_6=4 (-t-s-u) L_3$.

From the above relations we understand that $L_4$ has infinite singularities in u-channel and $-s L_6$ ($-t L_6$)has infinite t(s)-channel fermion poles and finally $-s L_3$ ($-t L_3$) has  double poles in $t, (t+s+u)$ and $s, (t+s+u)$   channels accordingly.

\vskip 0.1in

Therefore unlike the IIB amplitude , here we do have various double poles in different channels , which can not be explained by any duality transformations. Notice that if we apply momentum conservation, then we obtain
\beqa
t+s+u=-p^ap_a\non\eeqa

holds, thus this clearly shows that the expansion is low energy expansion and all  s,t,u  and $p^ap_a$ should be sent to zero, in order to be able to find out all  $\alpha'$ corrections.

 More importantly we obtained the complete form of the S-matrix in type IIA to be able to look for all the corrections to super Yang-Mills couplings and in particular remove the ambiguities related to the corrections by fixing the coefficients of new couplings. Once more from the above we can see that the second part of the amplitude is also antisymmetric with respect to fermions.

 Let us first deal with singularities and then get to the desired $\alpha'$ corrections of SYM and WZ couplings of type IIA superstring theory.

\section{ Infinite singularities for $<V_{C} V_{A} V_{\bar\psi} V_{\psi}>$ of IIA}

In \cite{Hatefi:2013eia} for the closed form of  $<V_{C} V_{\phi} V_{\bar\psi} V_{\psi}>$ in type IIB we have shown that there are infinite u-channel singularities for both massless gauge field and massless scalar field and in particular for the same S-matrix with different chirality , we have seen that there are only infinite massless scalar field of IIA. On the other hand, after having done further computations and provided physical kinematic reasons we confirmed that $<V_{C} V_{A} V_{\bar\psi} V_{\psi}>$ of type IIB proposed us that it just involves infinite massless scalar poles. Now for the same S-matrix (of course with different chirality) of IIA , here we have used various identities and observed that neither there are gauge nor scalar poles, therefore this clearly shows that the sum of the second term of ${\cal A}_{23}$, the fourth term of  ${\cal A}_{23}$ and all the terms appearing in ${\cal A}_{24}$ will be canceled off. Let us just provide the physical  reasons  in favor of showing there are no u-channel gauge /scalar poles left over.

\vskip.2in

As it is obvious from ${\cal A}_{23}$, the trace is just non zero for $p=n$ case and Ramond -Ramond ($C_{p-1}$) is allowed , or  $C_{p-1}\wedge F$ makes sense for this part of the S-matrix.
 on the other hand in order to have u- channel gauge poles in the field theory of IIA, we need to have Ramond -Ramond ($C_{p-3}$) and employ  $C_{p-3}\wedge F\wedge F$ , thus obviously  there is no any u-channel gauge poles.
Hence the following rule \footnote{
\beqa
{\cal A}&=& V_{\alpha}(C_{p-3},A_1,A)G_{\alpha\beta} (A) V_{\beta}(A,\bPsi_1,\Psi_2),\labell{amp5452}
\eeqa}
, and in particular the following corrections\footnote{\beqa
i\frac{\lambda^2\mu_p}{(p-2)!}\int d^{p+1}\sigma
 \,\sum_{n=-1}^{\infty}b_n(\alpha')^{n+1}\Tr\left(  C_{(p-3)} \wedge
D^{a_0}\cdots D^{a_n} F\wedge  D_{a_0}\cdots D_{a_n} F\right)\,
 \label{mm556}\eeqa} should not be taken into account for mixed RR, fermion /gauge of type IIA. Let us further discuss for not having u- channel scalar pole either. From ${\cal A}_{23}$, the trace is just non zero for $p=n$ case and Ramond -Ramond ($C_{p-1}$) is allowed and in particular to have u-channel scalar poles in the field theory of IIA, we need to have Ramond -Ramond ($C_{p-1}$) and consider $\partial_i C_{p-1}\wedge F \phi^i$\footnote{To work with  field theory of a RR and scalar fields, Wess-Zumino (WZ) terms, \cite{Myers:1999ps}, or  pull-back  or Taylor-expansion \cite{Hatefi:2012wj} are entirely explained.} while this is satisfactory, we should note that to have scalar poles in the closed form of the amplitude,  we need to have the kinematic factor of $\bar u \gamma^i u$ \footnote{\beqa
 {\cal A}   \sim \frac{\mu_p }{(p)!}
(\veps^v)^{a_0\cdots a_{p-2}ba}H^{i}_{a_0\cdots a_{p-2}}
  \alpha'\pi \xi_{1a}(2ik_{1b}) \bu_1^{\alpha}
 (\gamma_i) _{\alpha\beta} u_2^{\beta}\frac{1}{u}\Tr(\lam_1\lam_2\lam_3) \nonumber\eeqa

 } , which is not appeared in our S-matrix. Thus  unlike the closed form of
$<V_{C} V_{A} V_{\bar\psi} V_{\psi}>$ of IIB here we do not have any scalar u-channel poles. The above arguments hold even for
  ${\cal A}_{24}$, therefore we realize there are no gauge/ scalar poles for the closed form of $<V_{C} V_{A} V_{\bar\psi} V_{\psi}>$ of IIA.

  \vskip.2in

Hence the following rule \footnote{
\beqa
{\cal A}&=& V_{\alpha}^{i}(C_{p-1},A_1,\phi)G_{\alpha\beta}^{ij}(\phi) V_{\beta}^{j}(\phi,\bPsi_1,\Psi_2),\labell{amp5452}
\eeqa}
, more significantly the following corrections of type IIB  can not be held for type IIA.
\footnote{\beqa
i\frac{\lambda^2\mu_p}{p!}\int d^{p+1}\sigma
 \,\sum_{n=-1}^{\infty}b_n(\alpha')^{n+1}\Tr\left(\partial_i C_{(p-1)} \wedge
D^{a_0}\cdots D^{a_n}F D_{a_0}\cdots D_{a_n}\phi^i\right)\,
 \label{mm556}.\eeqa}

 \vskip.1in

 Let us now emphasize on infinite fermion poles.

 \vskip.1in

Having set the fact that total super ghost charge must be satisfied, we reveal that one can write down all fermion t-channel poles as follows:

\beqa
{\cal A}&\sim&  (P_{-}\fsH_{(n)}M_p)^{\alpha\beta}\xi_{1a} (2ik_{1b}) \bu^{\dga} u^{\dde}  \bigg\{\frac{-sL_6}{4(-s-u+\frac{1}{2})}\bigg[(\gamma^b \gab^a C)_\alpha{}^{\dga} C_\beta^{\dde}
+(\gamma^b \gab^a C)_\beta{}^{\dga}C_\alpha^{\dde}\bigg]\bigg\}\non\eeqa

where all s-channel fermion poles should be read off  as below\footnote{ where the normalization constant for this string amplitude should be  $\frac{\mu_p \pi^{-1/2}}{4}$.}

\beqa
{\cal A}&\sim&   (P_{-}\fsH_{(n)}M_p)^{\alpha\beta}\xi_{1a}(2ik_{1b}) \bu^{\dga} u^{\dde}  \bigg\{ \frac{tL_6}{4(-t-u+\frac{1}{2})}\bigg[(\gamma^b\gab^a C)_\alpha{}^{\dde}C_\beta^{\dga}+(\gamma^b \gab^a C)_\beta{}^{\dde}C_\alpha^{\dga}\bigg]\bigg\}
\non\eeqa

Obviously the amplitude is antisymmetric with respect to interchanging two fermions (or $t\leftrightarrow s$). Indeed the second term and the fourth term of $A_{21}$ ($A_{22}$ ) are due to an infinite number of $t$ ($s$) channel poles of mixed closed-open string of the type IIA amplitude.
Meanwhile the first and  the third term of $A_{21}$ ($A_{22}$ ) are related to different cases of $p,n$ and are due to an infinite number of contact interaction between one RR, two fermions with different chirality and one gauge field of type IIA.

Since the amplitude is antisymmetric, we just need to generate t-channel poles and finally by interchanging the fermions we could produce all infinite s-channel poles as well. If we include the expansion for $L_6$ and extract the related trace, then one can explore all infinite t-channel poles for $p=n$ case\footnote{
\beqa
{\cal A} &=& \frac{\mu_p \pi\xi_{1a}(\alpha') 2ik_{1b} }{p!} \bu^{\dga} (\ga^b)_{\dga\dde}  u^{\dde}  \sum_{n=-1}^{\infty}b_n  \frac{1}{t}(u+s)^{n+1}
(\veps)^{a_0\cdots a_{p-1}a}H_{a_0\cdots a_{p-1}}.
 \label{nmm1}\eeqa} and to be able to find them out, one needs to apply the following rule

\beqa
{\cal A}&=& V_{\alpha}(C_{p-1},\Psi_3,\bar\Psi)G_{\alpha\beta}(\Psi) V_{\beta}(\Psi,\bar\Psi_2,A_1),\labell{amp56099}
\eeqa
In \cite{Hatefi:2014saa} we have shown that neither fermions's kinetic term nor fermion propagator obtain any corrections \footnote{

 \beqa
 G_{\alpha\beta}(\psi) &=&\frac{-i\delta_{\alpha\beta}\ga^c(k_1+k_2)_c}{T_p t(2\pi\alpha') }.
 \nonumber\\
 V_{\beta}(\Psi,\bar\Psi_2,A_1)&=&-iT_p (2\pi\alpha')u^{\dde}\gamma^{a}_{\dde}\xi_{1a}\Tr(\lam_1\lam_2\lam^\beta)
 \label{mm5k5}\eeqa} where we have extracted the connection part in the kinetic term of fermions, now we need to supersymmetrize the Wess-Zumino actions to get involved fermions as follows:
 \footnote{ fermion field equations should be applied  $(\fsk_{2a}\bu=\fsk_{3a}u=0 )$ .}

 \beqa
\frac{(2\pi\alpha')\mu_p}{p!}\int d^{p+1}\sigma
\Tr\left( C_{a_0\cdots a_{p-2}} \bPsi \ga^c D_c \Psi \right) (\veps)^{a_0\cdots a_{p-2}}.\,
\label{nn2v2}\eeqa

Note that in above action we can just confirm the presence of partial derivative and in order to see whether or not the connection part kept fixed, one would need to go beyond five point functions, which definitely can not be answered by the computations of this paper, however,  in order to keep the general covariance we hold that. Now we would extract the vertex of one on-shell fermion/a RR $(p-1)$ form field and one off-shell fermion $V_{\alpha}(C_{p-1},\Psi_3,\bar\Psi)$\footnote{\beqa
 V_{\alpha}(C_{p-1},\Psi_3,\bar\Psi)&=&\frac{(2i\pi\alpha')\mu_p}{(p)!}
 H^c_{a_0\cdots a_{p-2}} \ga^c_{\dga} \bu^{\dga}
(\veps)^{a_0\cdots a_{p-2}}\,\Tr\left(\lambda_3\lambda^{\alpha}\right).\,
 \label{nn3m3}\eeqa}.

To discover all infinite fermion poles, one needs to actually have all order corrections of the above coupling, for which surprisingly these corrections have been found in type IIB  \cite{Hatefi:2013hca} and do work in IIA.\footnote{
\beqa
\frac{(2i\pi\alpha')\mu_p}{(p)!}\int d^{p+1}\sigma
\sum_{n=-1}^{\infty}b_n(\alpha')^{n+1}\Tr\left( C_{a_0\cdots a_{p-2}} D^{a_0}\cdots D^{a_n} \bPsi \ga^c D_{a_0}\cdots D_{a_n} D_c \Psi \right)\,(\veps)^{a_0\cdots a_{p-2}}.
\label{mnm}\eeqa}

Concerning \reef{mnm}, one finds the entire and all order $\alpha'$ corrections of $V_{\alpha}(C_{p-1},\Psi_3,\bar\Psi)$ as follows

 \beqa
 V_{\alpha}(C_{p-1},\Psi_3,\bar\Psi)=\frac{(2i\pi\alpha')\mu_p H^c_{a_0\cdots a_{p-2}} \ga^c_{\dga}}{(p)!}
 \bu^{\dga}
(\veps)^{a_0\cdots a_{p-2}}\,\Tr\left(\lambda_3\lambda^{\alpha}\right)\,\sum_{n=-1}^{\infty}b_n \bigg(\alpha'(k_3.k_2+k_1.k_3)\bigg)^{n+1}
 \label{kok}\eeqa

If we replace \reef{kok} and \reef{mm5k5} inside \reef{amp56099} and use fermion 'equations of motion , most notably we apply momentum conservation just in world volume direction $(k_1+k_2+k_3+p)^a=0$ (in transverse direction it is arbitrary) then we are precisely able to re-construct all orders in $\alpha'$ t(s)- channel fermion poles. Hence, we were able to clarify the fact RR $p-1$ form field has proposed the same $\alpha'$ corrections to two fermion fields of IIA.

\vskip.1in

We just highlight the point that these corrections are the key ingredient tools for all higher point functions such that by applying them , we could come over either the full world-sheet integrals of the higher point functions or find out all their singularities and therefore resolve the singularities of the string theory.

\vskip.2in

Let us end this section by producing all order  $\alpha'$ higher derivative corrections of a RR $p-1$ form field , two fermions of type IIA and a gauge field that would correspond to contact interactions extracted out of expansions  where we just write down the final result.\footnote{\beqa
&&\sum_{p,n,m=0}^{\infty}e_{p,n,m} (\alpha')^{2n+m-2}(\frac{\alpha'}{2})^{p}\frac{(2\pi\alpha')^2\mu_p}
{\pi p!}\int d^{p+1}\sigma
 \veps^{a_0\cdots a_{p-2}} \Tr \bigg( C_{a_0\cdots a_{p-2}}  \nonumber\\&&\times (D^aD_a)^p D^{a_1}\cdots D^{a_m}( D_{a_1}\cdots D_{a_n} A_c D_{a_{n+1}}\cdots D_{a_{2n}}\bPsi)   \ga^c  D_{a_1}\cdots D_{a_m}( D^{a_1}\cdots D^{a_n} D^{a_{n+1}}\cdots D^{a_{2n}}  \Psi )\bigg)
 \non\eeqa
 where the connection parts could be confirmed , if we would have had the result of higher point functions.} Indeed these contact interactions are produced by comparison with string theory S-matrix of type IIA.

\subsection{ New Wess-Zumino couplings of IIA and their   $\alpha'^3$ corrections  }

The aim of this section is to actually come over  double poles, new  Wess-Zumino couplings of IIA and their   $\alpha'^3$ corrections.
 Unlike the closed form of the correlation functions for various $p,n$ cases of one RR, two fermions and a gauge field of type IIB , here in IIA we discover some double poles and we are able to advance our knowledge about the closed form of SYM/WZ couplings. On the other hand we are going to emphasize within detail  that the following fact holds.

 \vskip.1in

 Although the general structure at third order of $\alpha' $ for the corrections to two on-shell fermions and one off-shell scalar and one on-shell gauge field (SYM) of IIA is the same, however the coefficient of those corrections are completely different from type IIA, which means that one has to first derive the S-matrix of type IIA and then by comparison those elements to field theory side, the coefficient of the field theory couplings should have been fixed without any ambiguity. We explain it further.

 \vskip.1in

 In order to get to the above conclusion, let us first write down the S-matrix of type IIA. Hence, we need to consider the first and the third term of ${\cal A}_{23} $ and in particular consider the world volume component of $\lambda=a$ in both terms of  ${\cal A}_{25} $, finally we need to add them up and extract the related traces, such that after having considered all the simplifications , one could explore the S-matrix as below:

   \beqa {\cal A} &=& \frac{ \alpha'^3ik_{1b}\pi^{-1/2}\mu_p}{p!} \veps^{a_0\cdots a_{p-1}b} H_{a_0\cdots a_{p-1}}\xi_{1a}\bu^{\dga} (\ga^a )_{\dga\dde} u^{\dde} (t) L_3
\Tr\left(\lambda_1\lambda_2\lambda_3\right)\,
 \labell{ampcmm},
\eeqa

 Note that we have the same S-matrix for $-s L_3$ but the terms inside that S-matrix can be precisely generated by interchanging the fermions, most importantly since the amplitude is antisymmetric under the interchanging of the fermions, we conclude this part has to be vanished for IIB with the same chirality, which is consistent with our expectation of IIA.

  \vskip.1in

The expansion of $t L_3$ was given but for the completeness we present it in the footnote\footnote{\beqa
tL_6&=&\frac{\sqrt{\pi}}{2}\left(\frac{-1}{s(t+s+u)}+ \frac{4\ln(2)}{s}+\left(\frac{\pi^2}{6}-8\ln(2)^2\right)\frac{(s+t+u)}{s}
+\cdots\right)\labell{high2}\eeqa}.

As we can see from the expansion we have a double pole in $s,(s+t+u)-$ channels ( there is no double pole in type IIB). If we make use of type IIA  effective field theory and RR's kinematic relations, then we can discover that the following rule is being held in IIA.

 \vskip.1in

\beqa
{\cal A}&=&V_a(C_{p-1},A)G_{ab}(A)V_b(A,\bar\psi_1,\psi_2)G(\psi_2)V(\bar\psi_2,\psi_1,A^{(1)})\labell{ampk1l5}\eeqa

where the vertex of one RR and an Abelian gauge field should have been read from the Chern-Simons term  \footnote{\beqa
(2\pi\alpha')\mu_p\int d^{p+1}\sigma {1\over p!}
\Tr\bigg(C_{p-1} \wedge F\bigg)\label{cvb}\eeqa} in which it has been fixed and there is no correction to that coupling so that $V_a(C_{p-1},A)=
(2\pi\alpha')\mu_p {1\over p!}
(\veps)^{a_0\cdots a_{p}}
H_{a_0\cdots a_{p}}$, meanwhile the gauge propagator must be derived from the kinetic term of gauge field which has no correction either  \footnote{
\beqa
G_{\alpha\beta}^{ab}(A) &=&\frac{-i\delta_{\alpha\beta}\delta^{ab}}{T_p(2\pi\alpha')^2
(t+s+u)} \label{gpop}\eeqa}.

 Now one may argue that the vertex of an on-shell/ an off-shell fermion and one off-shell gauge field has no contribution to this S-matrix. Because we need to extract the covariant derivative of fermion fields , however, notice that these fermions originated from two different D-branes so the following vertex makes sense in the presence of different overlapped  branes (note that similar couplings have been discovered for non super symmetric cases in brane anti brane systems ) so that :

\beqa
V_b(A,\bar\psi_1,\psi_2)&=& T_p (2\pi\alpha')\bar u^{\dga} \gamma_{b\dga}\label{dds}\eeqa
It is also discussed that the kinetic term of fermion has no correction. Concerning momentum conservation, one explores the fermion propagator as well as  $V(\bar\psi_2,\psi_1,A^{(1)})$.
\footnote{\beqa
G(\psi) &=&\frac{-i\gamma^a (k_1+k_3)_a}{T_p(2\pi\alpha')s}\nonumber\\
V(\bar\psi_2,\psi_1,A^{(1)})&=&T_p (2\pi\alpha') u^{\dde} \gamma_{\dde}^c \xi_{1c}\label{mko}\eeqa}

Now if we substitute \reef{gpop},\reef{dds}, \reef{mko} and the above Chern-Simons term inside the above rule \reef{ampk1l5}, we obtain

\vskip.2in

   \beqa {\cal A} &=& \frac{ 4ik_{1b} \mu_p}{p! (s) (t+s+u)} \veps^{a_0\cdots a_{p-1}b} H_{a_0\cdots a_{p-1}}\xi_{1a}\bu^{\dga} (\ga^a )_{\dga\dde} u^{\dde}  \Tr\left(\lambda_1\lambda_2\lambda_3\right)\,
 \labell{ampcmm3},
\eeqa

Hence we have generated the first double pole of the type IIA, where fermion field equations of motion have been applied to the above field theory side as well.

The natural question, one may ask is how to deal with the second term of the expansion, which is a simple s- channel pole. Replacing it to the S-matrix one might explore the second term of the S-matrix as

   \beqa {\cal A} &=& \frac{ ln(2) \alpha'^3ik_{1b}\mu_p }{2s p!} \veps^{a_0\cdots a_{p-1}b} H_{a_0\cdots a_{p-1}}\xi_{1a}\bu^{\dga} (\ga^a )_{\dga\dde} u^{\dde}
\Tr\left(\lambda_1\lambda_2\lambda_3\right)\,
 \labell{12ampcmm},
\eeqa

Let us begin with generating new WZ couplings as follows. In order to actually produce the above S-matrix , one has to take into account the following rule
\beqa
{\cal A}&=&V(C_{p-1},\bar\Psi_1,\Psi_2)G(\Psi_2)V(\bar\Psi_2,\Psi_1,A^1)\label{mdf}\eeqa

with new WZ coupling as
\beqa
\frac{(2i\pi\alpha')\mu_p}{p!}\beta_1^2
\Tr\left( C_{a_0\cdots a_{p-2}} \bPsi_1 \ga^b D_b \Psi_2 \right) (\veps)^{a_0\cdots a_{p-2}}
\label{lopm}\eeqa

whereas the fermion propagator and the structure of $V(\bar\Psi_2,\Psi_1,A^1)$ kept fixed. \footnote{ We should drop out the connection part inside the above WZ coupling \reef{lopm}}. Therefore , if we extract the needed vertex from \reef{lopm}  as

\beqa
\frac{(2i\pi\alpha')\mu_p}{p!}\beta_1^2
H_{a_0\cdots a_{p-1}} \bu^{\dga} (\ga^b )_{\dga}  (\veps)^{a_0\cdots a_{p-1}b}
\label{lopm1}\eeqa
with $\beta_{1}=(2ln2/(\pi\alpha'))^{1/2}$ and replace the given vertices appearing in \reef{mdf}, we will precisely obtain \reef{12ampcmm} and this means that not only can we confirm the new form of WZ coupling but also we fixed its normalization constant. One might wonder about the third simple pole, because it is again simple s-channel pole and we have already produced it. However the answer is hidden in the higher derivative correction of new WZ coupling that we just discovered right now.

\vskip.1in

 Indeed the above  rule $
{\cal A}=V(C_{p-1},\bar\Psi_1,\Psi_2)G(\Psi_2)V(\bar\Psi_2,\Psi_1,A^1) $ with the same fermion propagator,  likewise $V(\bar\Psi_2,\Psi_1,A^1)$ should be applied for the third term of the expansion, while the vertex $V(C_{p-1},\bar\Psi_1,\Psi_2)$  gets modified. Let us propose the following higher derivative correction  to a $(p-1)$-RR form field and one off-shell/on-shell fermion field as

 \beqa
\bigg(\frac{\pi^2}{6}-8 ln2^2\bigg)\frac{i (\alpha')^2\mu_p}{p!}
\Tr\left( C_{a_0\cdots a_{p-2}} D^aD_a (\bPsi_1 \ga^b D_b \Psi_2) \right) (\veps)^{a_0\cdots a_{p-2}}
\label{n1n2xx}\eeqa

 and take momentum conservation for all external states in the world volume direction   $s+t+u=-p^ap_a$ such that,

 \beqa
 V(C_{p-1},\bar\Psi_1,\Psi_2)&=& \alpha'^2 \mu_p \bigg(\frac{\pi^2}{6}-8 ln2^2\bigg) \frac{1}{p!}H_{a_0\cdots a_{p-1}} \bu^{\dga} (\ga^b )_{\dga} (t+s+u)(\veps)^{a_0\cdots a_{p-1}b}\label{mmn}\eeqa

Having replaced \reef{mmn} and the same vertex/propagator inside \reef{mdf}, we get to

 \beqa {\cal A} &=& \bigg(\frac{\pi^2}{6}-8 ln2^2\bigg)\frac{ (t+s+u)\xi_{1a} ik_{1b}\mu_p}{ s p!} (\veps)^{a_0\cdots a_{p-1}b} H_{a_0\cdots a_{p-1}}\bu^{\dga} (\ga^a )_{\dga\dde} u^{\dde}
\Tr\left(\lambda_1\lambda_2\lambda_3\right)\,
 \labell{ampbcgh},
\eeqa
 we have also used fermion field $k_{3a} \gamma^a u=0$ , thus the third pole is also precisely produced. In order to show that the coefficients of the higher derivative corrections to  SYM couplings of two fermion fields, one off-shell scalar and one on-shell gauge of type IIA are different from the type IIB ones, we just consider the last two terms of the first part of the amplitude  as:

 \beqa
{\cal A} & \sim& (P_{-}\fsH_{(n)}M_p)^{\alpha\beta}\xi_{1a} \bu ^{\dga} u ^{\dde}
  [ik_2^a s-ik_3^at](\gamma^\mu C)_{\alpha \beta}(\bar\gamma_\mu C)_{\dga\dde}L_3 \Tr(\lam_1\lam_2\lam_3)
\labell{amp3qpo},\eeqa

 If we consider the sum of the fourth terms of the expansions of $sL_3$ and $-t L_3$ in the above S-matrix,
\footnote{\beqa
sL_3&=&\frac{\sqrt{\pi}}{2}\left(\frac{-1}{t(t+s+u)}+ \frac{4\ln(2)}{t}+\left(\frac{\pi^2}{6}-8\ln(2)^2\right)\frac{(s+t+u)}{t}-\frac{\pi^2}{3}\frac{s}{(t+s+u)}+\cdots\right)\\
-tL_3&=&\frac{\sqrt{\pi}}{2}\left(\frac{1}{s(t+s+u)}- \frac{4\ln(2)}{s}-\left(\frac{\pi^2}{6}-8\ln(2)^2\right)\frac{(s+t+u)}{s}+\frac{\pi^2}{3}\frac{t}{(t+s+u)}+\cdots\right)
\labell{high2}\\\eeqa} extract the trace for the transverse component of $(\mu=i)$ and for $p+2=n$ case with a proper normalization constant we derive the S-matrix as follows:

 \beqa {\cal A} &=& \frac{ i\alpha'^3\pi^2\mu_p}{12(t+s+u)(p+1)!} (\veps)^{a_0\cdots a_{p}} H^i_{a_0\cdots a_{p}} \bu^{\dga} (\ga^i )_{\dga\dde} u^{\dde} (-sk_2.\xi_1+tk_3.\xi_1)
\Tr\left(\lambda_1\lambda_2\lambda_3\right)
 \labell{ampcmmgyd}
\eeqa

This S-matrix with three momenta ( without counting RR 's momentum) has to be produced by the following rule of the effective field theory

\beqa
A&=&   V^{\alpha}_i(C_{p+1},\phi) G^{\alpha\beta} _{ij} (\phi) V^{\beta}_{j}(\phi,\bar\Psi,\Psi,A_1)\label{esi34b}\eeqa

The RR's kinetic term has proposed a constraint that just massless scalar can be propagated to generate the  $(t+s+u)$- pole. Hence one might wonder whether or not the couplings that we found for two fermions and one scalar/gauge of type IIB \footnote{ \beqa
&&{\cal L}^{n,m}= \pi^3\alpha'^{n+m+3}T_p\bigg(a_{n,m}\Tr\bigg[\cD_{nm}\left(\bPsi \ga^i D_b\Psi D^a\phi^i F_{ab} \right)+\cD_{nm}\left( \bPsi \ga^i D_b\Psi F_{ab} D^a\phi^i \right)
\nonumber\\&&+h.c \bigg]
+i b_{n,m}\Tr\bigg[\cD'_{nm}\left(\bPsi \ga^i D_b\Psi D^a\phi^i F_{ab} \right)+\cD'_{nm}\left(
 \bPsi \ga^i D_b\Psi F_{ab} D^a\phi^i  \right)+h.c.\bigg]
\bigg).
\non
\eeqa } even at third order in $\alpha', n=m=0$  could be applied to IIA.

Let us first consider the third order corrections of two on-shell fermions, one off-shell scalar and one on-shell gauge of type IIB . \footnote{\beqa
\frac{T_p (2\pi\alpha')^3}{4}\bigg[\bPsi \ga^i D_b\Psi D^a\phi_i F_{ab}+\bPsi \ga^i D_b\Psi F_{ab} D^a\phi_i\bigg] \label{zzxc2}
 \eeqa}

The scalar propagator should be derived from its kinetic term and the vertex of one RR $(p+1)-$ form field and an off-shell scalar is given in \cite{Hatefi:2013eia} but for the completeness we mention both once more. \footnote{\beqa
G_{\alpha\beta}^{ij}(\phi) &=&\frac{-i\delta_{\alpha\beta}\delta^{ij}}{T_p(2\pi\alpha')^2
k^2}=\frac{-i\delta_{\alpha\beta}\delta^{ij}}{T_p(2\pi\alpha')^2
(t+s+u)},\nonumber\\
V_{\alpha}^{i}(C_{p+1},\phi)&=&i(2\pi\alpha')\mu_p\frac{1}{(p+1)!}(\veps)^{a_0\cdots a_{p}}
 H^{i}_{a_0\cdots a_{p}}\Tr(\lambda_{\alpha}).\labell{Fey}\eeqa} Taking  \reef{zzxc2} and  considering  the orderings of $\Tr(\lambda_2\lambda_3\lambda_{\beta}\lambda_1)$ and $\Tr(\lambda_2\lambda_3\lambda_1\lambda_{\beta})$ for the first and second couplings of \reef{zzxc2} accordingly, we are able to get to

\beqa
V_{\beta}^{j}(\phi,\bar\Psi,\Psi,A_1)&=& -i\frac{T_p (2\pi\alpha')^3}{4}
\bu^{\dga} (\ga^j )_{\dga\dde} u^{\dde} \bigg( -tk_3.\xi_1+s k_2.\xi_1\bigg)\Tr(\lam_1\lam_2\lam_3\lam_\beta)
\label{eesddem1}\eeqa

where all the connection parts or commutator terms should have been ignored in the those corrections and momentum conservation has to be used.
\vskip.1in

If we would replace  \reef{eesddem1}  and  \reef{Fey} inside the rule of \reef{esi34b} we would obtain the following field theory amplitude of type IIA

 \beqa {\cal A} &=& \frac{ -i\alpha'^3\pi^2\mu_p}{2(t+s+u)(p+1)!} (\veps)^{a_0\cdots a_{p}} H^j_{a_0\cdots a_{p}} \bu^{\dga} (\ga^j )_{\dga\dde} u^{\dde} (sk_2.\xi_1-tk_3.\xi_1)
\Tr\left(\lambda_1\lambda_2\lambda_3\right)
 \labell{ampcmmgyd65}
\eeqa

Its comparison with string theory S-matrix of \reef{ampcmmgyd}, shows us that obviously the coefficients of the corrections of type IIB must be modified (to make sense of IIA computations) in such a way that we can generate all the string theory S-matrix elements of type IIA very precisely.

\vskip.2in

The proposal for SYM corrections of type IIA is as follows:

\beqa
\frac{T_p (\pi\alpha')^3}{3}\bigg[\bPsi \ga^i D_b\Psi D^a\phi_i F_{ab}+\bPsi \ga^i D_b\Psi F_{ab} D^a\phi_i\bigg] \label{zzxc27}
 \eeqa

The scalar propagator and the vertex of one RR $(p+1)-$ form field and an off-shell scalar remain invariant, now taking  into account \reef{zzxc27} and  considering  the orderings of $\Tr(\lambda_2\lambda_3\lambda_{\beta}\lambda_1)$ and $\Tr(\lambda_2\lambda_3\lambda_1\lambda_{\beta})$ for the first and second couplings of \reef{zzxc27} accordingly, we are able to achieve

\beqa
V_{\beta}^{j}(\phi,\bar\Psi,\Psi,A_1)&=& -i\frac{T_p (\pi\alpha')^3}{3}
\bu^{\dga} (\ga^j )_{\dga\dde} u^{\dde} \bigg( -tk_3.\xi_1+s k_2.\xi_1\bigg)\Tr(\lam_1\lam_2\lam_3\lam_\beta)
\label{eesddem15}\eeqa

where all the connection parts or commutator terms are removed in the those corrections and momentum conservation was used.
\vskip.1in

If we would replace  \reef{eesddem15}  and \reef{Fey} inside the rule of \reef{esi34b} we would find out the following field theory amplitude of type IIA

 \beqa {\cal A} &=& \frac{ -i\alpha'^3\pi^2\mu_p}{12(t+s+u)(p+1)!} (\veps)^{a_0\cdots a_{p}} H^j_{a_0\cdots a_{p}} \bu^{\dga} (\ga^j )_{\dga\dde} u^{\dde} (sk_2.\xi_1-tk_3.\xi_1)
\Tr\left(\lambda_1\lambda_2\lambda_3\right)
 \labell{ampcmmgyd65}
\eeqa

which is precisely the  string theory S-matrix of \reef{ampcmmgyd} , this dictates that we have correctly derived the third corrections of SYM to type IIA. It is shown that all order $\alpha'$ corrections of type IIB for fermionic amplitudes and bosonic amplitudes could be derived by applying a universal prescription to their leading order couplings \cite{Hatefi:2012rx} (see \cite{Hatefi:2010ik,Hatefi:2012ve}), however, it is not obvious that kind of rule holds for type IIA.

\vskip.2in

Let us end this section by talking about some other couplings that one could expect to have seen them in type IIA. Namely for $p=n$ case , one might  hope to obtain the  other SYM couplings of two on-shell fermions and one off-shell/ an on-shell gauge then fix their coefficients in both type IIA and IIB.  First of all one has to write down all possible couplings with three momenta  as follows:

\beqa
\bPsi \ga^a D_a\Psi F_{bc} F^{bc}\quad \bPsi \ga^a D_b\Psi F_{ac} F^{bc}  \quad \bPsi \ga^a D_b\Psi F_{bc} F^{ac}\ \label{UUU}\eeqa
If we take two different orderings of $\Tr(\lambda_2\lambda_3\lambda_1\lambda_{\beta})$ and $\Tr(\lambda_2\lambda_3\lambda_{\beta}\lambda_1)$ with Abelian gauge field $(\lambda_{\beta})$ , then we conclude that the contribution of the second and third term to effective field theory is the same  so we just need to keep track of one of them. On the other hand the first coupling produces the following terms  \footnote{\beqa
v_{b}^{\beta}(\bar\Psi_2,\Psi_3,A_1,A)&=&\bar u \ga^a u (-ik_{3a}) \bigg[-(t+s) \xi^{b}  +2k^{b}_1(-k_2.\xi_1-k_3.\xi_1)\bigg] \Tr(\lambda_1\lambda_2\lambda_3\lambda_{\beta}) \label{nml}\eeqa} where having taken on-shell condition for fermion field, we realize that the first coupling of \reef{UUU} does not play any rule in the effective field theory of this S-matrix. What about the the second coupling of \reef{UUU}.  If we consider that coupling, apply momentum conservation, take all possible orderings and re-consider the on-shell condition for fermions we obtain the following terms:

\beqa
v_{c}^{\beta}(\bar\Psi_2,\Psi_3,A_1,A)&=& (-i)\bar u \ga^a u \bigg[k_{1a} (\frac{t}{2}\xi_{1c}-(k_2.\xi_1+k_3.\xi_1)k_{3c})-\frac{t}{2} \xi_{1a}k_{1c}-(\frac{t}{2}+\frac{s}{2})\xi_{1a}k_{3c}\nonumber\\&&+k_{1a} (-\frac{s}{2}\xi_{1c}-k_3.\xi_1 k_{1c})+\frac{s}{2} k_2.\xi_1-\frac{t}{2} k_3.\xi_1\bigg] \Tr(\lambda_1\lambda_2\lambda_3\lambda_{\beta}) \label{nml}\eeqa

for which these terms are not appeared in the S-matrix elements of one RR, two fermion fields (with different chirality) and one gauge field of IIA.  Therefore we can definitely conclude that there exist no even first order of $\alpha'$ correction to those SYM couplings of type IIA.

\section{Concluding remarks }

In this paper,  first we obtained the  compact/closed form of the correlation function of four spin operators with different chirality and one current  $<V_{C}V_{\bar\psi}^{\dga}V_{\psi}^{\dde} V_{A}>$ in ten dimensions of space time in type IIA superstring theory.  Conformal  field theory techniques have been  used to explore the entire form of the S-matrix of one closed string RR, two fermion fields with different chirality and one gauge field of IIA. Concerning the complete form of IIA S-matrix, we are able to first of all find out the third order of $\alpha'$ higher derivative corrections to SYM couplings and secondly distinguish $\alpha'$ corrections from type IIB. Our computations show that the coefficients of two fermions (with different chirality), one scalar and one gauge of IIA  corrections should be completely different from type IIB while in  \cite{Hatefi:2014saa}  we have proven not only the coefficients of two fermions- two scalars but also the the general structure of those corrections of IIA is entirely different from IIB corrections at all order of $\alpha'$.

\vskip.2in

The other result that we obtained is that, although in the S-matrix we have u-channel pole but if we use various identities those  scalar/gauge singularities will be disappeared so  unlike the closed form of $<V_{C}V_{\bar\psi} V_{\psi} V_{A}>$ of type IIB we have no u-channel singularity in IIA. Note also that unlike the computations of IIB for the same S-matrix, in IIA we do have several double poles in $t,(t+s+u)$ and  in $s,(t+s+u)$ channels  where the presence of these double poles can not be confirmed by duality transformation at all. We have produced all infinite massless fermion poles of   $t,s$-channels  of type IIA for diverse $p,n$ cases. Most importantly  their all order $\alpha'$ higher derivative corrections and their extensions have been discovered and constructed without any ambiguities.

\vskip.2in

Having found the entire form of the S-matrix, we are able to discover several new Wess-Zumino couplings
with their third order $\alpha' $ corrections. Note that those new WZ couplings can not be found by having the S-matrix of IIB. It was just the first  $(t+s+u)-$ channel scalar pole, that dictated us that the $\alpha'^3$  corrections of IIA have to be accompanied with the same corrections but with different coefficients of IIB. We take this fact as a benefit of having done direct conformal  field theory techniques rather than applying duality transformation. By comparison field theory amplitude with string theory elements we also explored that  there is no even single  $\alpha'$ correction tow gauges/two fermions of type IIA.

\vskip 0.1in

Finally, we used CFT for propagators and as it is known the RR has two sectors and  it is not obvious how to get involved  its second sector $\tilde{\alpha}_n$ but one might apply the ideas appeared in  \cite{Billo:2006jm} for further developments so that clearly the analytic continuation must be made. In the other words background fields should be seen as functions of SYM. We could observe background fields as some composite states of the open strings where the certain definition of Taylor expanded function is necessary \cite{Myers:1999ps}.

\vspace{.3in}  
\section*{Acknowledgements}

I would like to thank J.Polchinski, E.Witten, N.Arkani-Hamed, N.Lambert, K.S Narain, P.Vanhove, J.Schwarz,P. Horava and F.Quevedo for very useful discussions. Some  stages of this work have been carried out  during my visits to Institute for advanced study in Princeton-NJ, University of California at Berkeley,CA, KITP, Harvard University, California Institute of Technology, 452-48, Pasadena, CA 91125, USA and at Simons Center for Geometry and Physics, Stony Brook University,Stony Brook, NY 11794, USA. The author also thanks  L.Alvarez-Gaume, I.Antoniadis and CERN theory division for the their hospitality.



\begin{thebibliography}{2007}

\bibitem{Polchinski:1995mt}
  J.~Polchinski,``Dirichlet-Branes and Ramond-Ramond Charges,''
  Phys.\ Rev.\ Lett.\  {\bf75}, 4724 (1995)
  [arXiv:hep-th/9510017].
\bibitem{Witten:1995im}
  E.~Witten,``Bound states of strings and p-branes,''
  Nucl.\ Phys.\  B {\bf 460},335 (1996)
  [arXiv:hep-th/9510135].
\bibitem{Polchinski:1996na}
  J.~Polchinski,`` Lectures on D-branes,''
  [arXiv:hep-th/9611050]
  ;
  C.~P.~Bachas,
  ``Lectures on D-branes,''
  [arXiv:hep-th/9806199].


\bibitem{Witten:2013tpa}
  E.~Witten,
   ``Notes On Holomorphic String And Superstring Theory Measures Of Low Genus,''
  arXiv:1306.3621 [hep-th].

\bibitem{Witten:2013cia}
  E.~Witten,
   ``More On Superstring Perturbation Theory,''
  arXiv:1304.2832 [hep-th].

\bibitem{D'Hoker:2013eea}
  E.~D'Hoker and M.~B.~Green,
   ``Zhang-Kawazumi Invariants and Superstring Amplitudes,''
  arXiv:1308.4597 [hep-th].

\bibitem{Bjerrum-Bohr:2014qwa}
  N.~E.~JBjerrum-Bohr, P.~H.~Damgaard, P.~Tourkine and P.~Vanhove,
   ``Scattering Equations and String Theory Amplitudes,''
  arXiv:1403.4553 [hep-th].

\bibitem{Arkani-Hamed:2013jha}
  N.~Arkani-Hamed and J.~Trnka,
   ``The Amplituhedron,''
  arXiv:1312.2007 [hep-th].

\bibitem{Arkani-Hamed:2013kca}
  N.~Arkani-Hamed and J.~Trnka,
   ``Into the Amplituhedron,''
  arXiv:1312.7878 [hep-th].

\bibitem{ArkaniHamed:2012nw}
  N.~Arkani-Hamed, J.~L.~Bourjaily, F.~Cachazo, A.~B.~Goncharov, A.~Postnikov and J.~Trnka,
  arXiv:1212.5605 [hep-th].





\bibitem{Ademollo:1974fc}
  M.~Ademollo, A.~D'Adda, R.~D'Auria, E.~Napolitano, P.~Di Vecchia, F.~Gliozzi and S.~Sciuto,
  ``Unified Dual Model for Interacting Open and Closed Strings,''
  Nucl.\ Phys.\ B {\bf 77}, 189 (1974).

\bibitem{Polchinski:1996nb}
  J.~Polchinski,
   ``String duality: A Colloquium,''
  Rev.\ Mod.\ Phys.\  {\bf 68}, 1245 (1996)
  [hep-th/9607050].

\bibitem{Hatefi:2012sy}
  E.~Hatefi, A.~J.~Nurmagambetov and I.~Y.~Park,
  ``$N^3$ entropy of $M5$ branes from dielectric effect,''
  Nucl.\ Phys.\ B {\bf 866}, 58 (2013)
  [arXiv:1204.2711 [hep-th]].


\bibitem{Hatefi:2012bp}
  E.~Hatefi, A.~J.~Nurmagambetov and I.~Y.~Park,
   ``ADM reduction of IIB on $\mathcal{H}^{p,q}$ to dS braneworld,''
  JHEP {\bf 1304}, 170 (2013)
  [arXiv:1210.3825 [hep-th]].
















\bibitem{Polchinski:1994fq}
 J.~Polchinski, ``Combinatorics of boundaries in string theory,''
Phys. Rev. {\bf D50},6041 (1994)[arXiv:hep-th/9407031]
;
  J.~Polchinski, S.~Chaudhuri and C.~V.~Johnson,
  ``Notes on D-Branes,''
  [arXiv:hep-th/9602052].
\bibitem{Hatefi:2014saa}
  E.~Hatefi,
   ``SYM, Chern-Simons, Wess-Zumino Couplings and their higher derivative corrections in IIA Superstring theory,''
  arXiv:1403.1238 [hep-th],Published in Eur.Phys.J. C74 (2014) 2949

\bibitem{Myers:1999ps}
  R.~C.~Myers,``Dielectric-branes,''
  JHEP {\bf 9912}, 022 (1999)
  [arXiv:hep-th/9910053].
\bibitem{Hatefi:2012zh}
  E.~Hatefi,
   ``Shedding light on new Wess-Zumino couplings with their corrections to all orders in alpha-prime,''
  JHEP {\bf 1304}, 070 (2013)
  [arXiv:1211.2413 [hep-th]].



\bibitem{Howe:2006rv}
  P.~S.~Howe, U.~Lindstrom and L.~Wulff,
   `On the covariance of the Dirac-Born-Infeld-Myers action,''
  JHEP {\bf 0702}, 070 (2007)
  [hep-th/0607156].
\bibitem{Leigh:1989jq}
  R.~G.~Leigh,
  ``Dirac-Born-Infeld Action from Dirichlet Sigma Model,''
  Mod.\ Phys.\ Lett.\ A {\bf 4}, 2767 (1989).
\bibitem{Cederwall:1996pv}
  M.~Cederwall, A.~von Gussich, B.~E.~W.~Nilsson and A.~Westerberg,
  ``The Dirichlet super three-brane in ten-dimensional type IIB supergravity,''
  Nucl.\ Phys.\ B {\bf 490}, 163 (1997)
  [hep-th/9610148]
  ;
  M.~Aganagic, C.~Popescu and J.~H.~Schwarz,
  ``D-brane actions with local kappa symmetry,''
  Phys.\ Lett.\ B {\bf 393}, 311 (1997)
  [hep-th/9610249]
;
  M.~Aganagic, C.~Popescu and J.~H.~Schwarz,
  ``Gauge invariant and gauge fixed D-brane actions,''
  Nucl.\ Phys.\ B {\bf 495}, 99 (1997)
  [hep-th/9612080]
  ;
  M.~Cederwall, A.~von Gussich, B.~E.~W.~Nilsson, P.~Sundell and A.~Westerberg,
  Nucl.\ Phys.\ B {\bf 490}, 179 (1997)
  [hep-th/9611159]
  ;
  E.~Bergshoeff and P.~K.~Townsend,
  Nucl.\ Phys.\ B {\bf 490} (1997) 145
  [hep-th/9611173].
\bibitem{Tseytlin:1999dj}
  A.~A.~Tseytlin,
   ``Born-Infeld action, supersymmetry and string theory,''
  In *Shifman, M.A. (ed.): The many faces of the superworld* 417-452
  [hep-th/9908105].

\bibitem{Tseytlin:1997csa}
  A.~A.~Tseytlin,
   ``On nonAbelian generalization of Born-Infeld action in string theory,''
  Nucl.\ Phys.\ B {\bf 501}, 41 (1997)
  [hep-th/9701125].


\bibitem{Hatefi:2010ik}
  E.~Hatefi,
  ``On effective actions of BPS branes and their higher derivative
  corrections,''
  JHEP {\bf 1005}, 080 (2010)
  [arXiv:1003.0314 [hep-th]].



\bibitem{Hatefi:2012wj}
  E.~Hatefi,
   ``On higher derivative corrections to Wess-Zumino and Tachyonic actions in type II super string theory,''
  Phys.\ Rev.\ D {\bf 86}, 046003 (2012)
  [arXiv:1203.1329 [hep-th]].

\bibitem{Hatefi:2013eia}
  E.~Hatefi,
  ``Closed string Ramond-Ramond proposed higher derivative interactions on fermionic amplitudes in IIB,''
  Nucl.\ Phys.\ B {\bf 880}, 1 (2014)
  [arXiv:1302.5024 [hep-th]].
\bibitem{Hatefi:2013yxa}
  E.~Hatefi,
  ``Selection Rules and RR Couplings on Non-BPS Branes,''
  JHEP {\bf 1311}, 204 (2013)
  [arXiv:1307.3520].
\bibitem{Hatefi:2013mwa}
  E.~Hatefi,
  ``All order $\alpha'$ higher derivative corrections to non-BPS branes of type IIB Super string theory,''
  JHEP {\bf 1307}, 002 (2013)
  [arXiv:1304.3711 [hep-th]].

\bibitem{Hatefi:2012cp}
  E.~Hatefi,
   ``On D-brane anti D-brane effective actions and their corrections to all orders in alpha-prime,''
  JCAP {\bf 1309}, 011 (2013)
  [arXiv:1211.5538 [hep-th]].

\bibitem{Garousi:2007fk}
  M.~R.~Garousi and E.~Hatefi,``On Wess-Zumino terms of Brane-Antibrane systems,''
  Nucl.\ Phys.\  B {\bf 800}, 502 (2008)
  [arXiv:0710.5875 [hep-th]].
\bibitem{Hatefi:2008ab}
  M.~R.~Garousi and E.~Hatefi,``More on WZ actions of non-BPS branes,''
  JHEP {\bf 0903}, 08 (2009)
  [arXiv:0812.4216 [hep-th]].



\bibitem{Koerber:2002zb}
  P.~Koerber and A.~Sevrin,
  ``The NonAbelian D-brane effective action through order alpha-prime**4,''
  JHEP {\bf 0210}, 046 (2002)
  [hep-th/0208044].

\bibitem{Keurentjes:2004tu}
  A.~Keurentjes, P.~Koerber, S.~Nevens, A.~Sevrin and A.~Wijns,
  ``Towards an effective action for D-branes,''
  Fortsch.\ Phys.\  {\bf 53}, 599 (2005)
  [hep-th/0412271].

\bibitem{Denef:2000rj}
  F.~Denef, A.~Sevrin and J.~Troost,
  ``NonAbelian Born-Infeld versus string theory,''
  Nucl.\ Phys.\ B {\bf 581}, 135 (2000)
  [hep-th/0002180].
\bibitem{Hashimoto:1996bf}
A.~Hashimoto and I.~R.~Klebanov,
``Scattering of strings from D-branes,''
Nucl.\ Phys.\ Proc.\ Suppl.\  {\bf 55B}, 118 (1997)
[arXiv:hep-th/9611214]
;
I.\ R.\ Klebanov and L.\ Thorlacius,
``The Size of p-branes,''
Phys. Lett.~ {\bf B371}, 51 (1996)[arXiv:hep-th/9510200]
 ;
S.S.\ Gubser, A.\ Hashimoto, I.R.\ Klebanov, and J.M.\ Maldacena,
``Gravitational lensing by $p$-branes,''Nucl.\ Phys. {\bf  B472}, 231 (1996) [arXiv:hep-th/9601057]
;
C.\ Bachas, ``D-Brane Dynamics,'' Phys. Lett.~ {\bf B374}, 37 (1996)[arXiv:hep-th/9511043]
;
  W.~Taylor,``Lectures on D-branes, gauge theory and M(atrices),''
  [arXiv:hep-th/9801182];
  C.~Vafa,``Lectures on strings and dualities,''
  [arXiv:hep-th/9702201]
;
  ;%
  M.~Billo, M.~Frau, F.~Lonegro and A.~Lerda,``N = 1/2 quiver gauge theories from open strings with R-R fluxes,''
  JHEP {\bf 0505},047 (2005)
  [arXiv:hep-th/0502084]
  ;
  M.~Billo, P.~Di Vecchia, M.~Frau, A.~Lerda, I.~Pesando, R.~Russo and S.~Sciuto,``Microscopic string analysis of the D0-D8 brane system and dual R-R
  states,''
  Nucl.\ Phys.\  B {\bf 526},199 (1998)
  [arXiv:hep-th/9802088].
  ;
  E.~Hatefi,
  ``Three Point Tree Level Amplitude in Superstring Theory,''
  Nucl.\ Phys.\ Proc.\ Suppl.\  {\bf 216}, 234 (2011)
  [arXiv:1102.5042 [hep-th]].

\bibitem{Hashimoto:1996kf}
  A.~Hashimoto and I.~R.~Klebanov,
   ``Decay of excited D-branes,''
  Phys.\ Lett.\ B {\bf 381}, 437 (1996)
  [hep-th/9604065].


\bibitem{Lambert:2003zr}
  N.~D.~Lambert, H.~Liu and J.~M.~Maldacena,
   ``Closed strings from decaying D-branes,''
  JHEP {\bf 0703}, 014 (2007)
  [hep-th/0303139].

\bibitem{Dudas:2001wd}
  E.~Dudas, J.~Mourad and A.~Sagnotti,
   ``Charged and uncharged D-branes in various string theories,''
  Nucl.\ Phys.\ B {\bf 620}, 109 (2002)
  [hep-th/0107081].

\bibitem{Antoniadis:1999xk}
  I.~Antoniadis, E.~Dudas and A.~Sagnotti,
   ``Brane supersymmetry breaking,''
  Phys.\ Lett.\ B {\bf 464}, 38 (1999)
  [hep-th/9908023].

\bibitem{deAlwis:2013gka}
  S.~de Alwis, R.~Gupta, E.~Hatefi and F.~Quevedo,
  ``Stability, Tunneling and Flux Changing de Sitter Transitions in the Large Volume String Scenario,''
  JHEP {\bf 1311}, 179 (2013)
  [arXiv:1308.1222 [hep-th]].










\bibitem{Ferrari:2013pi}
  F.~Ferrari,
   ``On Matrix Geometry and Effective Actions,''
  Nucl.\ Phys.\ B {\bf 871}, 181 (2013)
  [arXiv:1301.3722 [hep-th]].
\bibitem{Hatefi:2012ve}
  E.~Hatefi and I.~Y.~Park,
  ``More on closed string induced higher derivative interactions on D-branes,''
  Phys.\ Rev.\ D {\bf 85}, 125039 (2012)
  [arXiv:1203.5553 [hep-th]].
\bibitem{Hatefi:2012rx}
  E.~Hatefi and I.~Y.~Park,
  ``Universality in all-order $\alpha'$ corrections to BPS/non-BPS brane world volume theories,''
  Nucl.\ Phys.\ B {\bf 864} (2012) 640
  [arXiv:1205.5079 [hep-th]].


\bibitem{Boels:2010bv}
  R.~H.~Boels, D.~Marmiroli and N.~A.~Obers,
  ``On-shell Recursion in String Theory,''
  JHEP {\bf 1010}, 034 (2010)
  [arXiv:1002.5029 [hep-th]].
\bibitem{Maxfield:2013wka}
  T.~Maxfield, J.~McOrist, D.~Robbins and S.~Sethi,
  ``New Examples of Flux Vacua,''
  arXiv:1309.2577 [hep-th].
\bibitem{McOrist:2012yc}
  J.~McOrist and S.~Sethi,
   ``M-theory and Type IIA Flux Compactifications,''
  JHEP {\bf 1212}, 122 (2012)
  [arXiv:1208.0261 [hep-th]];
  E.~Hatefi, A.~J.~Nurmagambetov and I.~Y.~Park,
   ``Near-Extremal Black-Branes with n*3 Entropy Growth,''
  Int.\ J.\ Mod.\ Phys.\ A {\bf 27}, 1250182 (2012)
  [arXiv:1204.6303 [hep-th]].

\bibitem{Vafa:1996xn}
  C.~Vafa,
   ``Evidence for F theory,''
  Nucl.\ Phys.\ B {\bf 469}, 403 (1996)
  [hep-th/9602022].


\bibitem{Park:2007mc}
I.~Y.~Park,``Open string engineering of D-brane geometry,''
  JHEP {\bf 0808}, 026 (2008)
  [arXiv:0806.3330[hep-th]].


\bibitem{Bilal:2001hb}
  A.~Bilal,
   ``Higher derivative corrections to the nonAbelian Born-Infeld action,''
  Nucl.\ Phys.\ B {\bf 618}, 21 (2001)
  [hep-th/0106062].

\bibitem{Barreiro:2012aw}
  L.~A.~Barreiro and R.~Medina,
   ``Revisiting the S-matrix approach to the open superstring low energy effective lagrangian,''
  JHEP {\bf 1210}, 108 (2012)
  [arXiv:1208.6066 [hep-th]].


\bibitem{Barreiro:2013dpa}
  L.~A.~Barreiro and R.~Medina,
   ``RNS derivation of N-point disk amplitudes from the revisited S-matrix approach,''
  arXiv:1310.5942 [hep-th].
\bibitem{Kennedy:1999nn}
  C.~Kennedy and A.~Wilkins,``Ramond-Ramond couplings on brane-antibrane systems,''
  Phys.\ Lett.\  B {\bf 464}, 206 (1999)
  [arXiv:hep-th/9905195].

 \bibitem{Chandia:2003sh}
  L.~A.~Barreiro and R.~Medina,``5-field terms in the open superstring effective action,''
  JHEP {\bf 0503},055 (2005)
  [arXiv:hep-th/0503182].
;
  R.~Medina, F.~T.~Brandt and F.~R.~Machado,``The open superstring 5-point amplitude revisited,''
  JHEP {\bf 0207},071 (2002)
  [arXiv:hep-th/0208121]






\bibitem{Hatefi:2013hca}
  E.~Hatefi,
   ``SuperYang-Mills, Chern-Simons couplings and their all order $\alpha'$ corrections in IIB superstring theory,''
  arXiv:1310.8308 [hep-th],Published in Eur.Phys.J. C74 (2014) 8, 3003.



\bibitem{Polchinski:1998aa}
J.~Polchinski,``String theory '',Vol 2,Cambridge University Press,1998





\bibitem{Friedan:1985ge}
  D.~Friedan, E.~J.~Martinec and S.~H.~Shenker,
  ``Conformal Invariance, Supersymmetry and String Theory,''
  Nucl.\ Phys.\ B {\bf 271}, 93 (1986).



\bibitem{Fotopoulos:2001pt}
  A.~Fotopoulos,``On (alpha')**2 corrections to the D-brane action for non-geodesic
  world-volume embeddings,''
  JHEP {\bf 0109}, 005 (2001)
  [arXiv:hep-th/0104146].


\bibitem{Billo:2006jm}
  M.~Billo, M.~Frau, F.~Fucito and A.~Lerda,
  ``Instanton calculus in R-R background and the topological string,''
  JHEP {\bf 0611}, 012 (2006)
  [hep-th/0606013].

















%



\end{thebibliography}
\end{document}